\documentclass{jkas}

\def\year{2016} 
\def\month{September} 
\def\volume{00} 
\def\issue{0} 
\def\beginpage{1} 
\def\endpage{99} 
\def\received{---} 
\def\accepted{---} 

\date{Received \received ; accepted \accepted}

\newcommand{\bdv}[1]{\mbox{\boldmath$#1$}}

\def\masyr{{\rm mas}\,{\rm yr}^{-1}}
\def\muas{{\mu\rm as}}

\def\min{{\rm min}}
\def\mas{{\rm mas}}
\def\lim{{\rm lim}}

\def\bOmega{{\bdv\Omega}}

\def\apj{{ApJ}}
\def\aj{{AJ}}

\def\jkas{{JKAS}}

\def\mnras{{MNRAS}}
\def\yr{{\rm yr}}

\title{
     Predicted Information Content of Gaia-Hipparcos
}

\author[1,2,3]{Andrew Gould}
\author[4]{Juna A. Kollmeier}

\affil[1]{Max-Planck-Institute for Astronomy, K\"onigstuhl 17, 69117 Heidelberg, Germany; \email{gould@astronomy.ohio-state.edu }}
\affil[2]{Korea Astronomy and Space Science Institute, Daejon 305-348, Republic of Korea}
\affil[3]{Department of Astronomy Ohio State University, 140 W.\ 18th Ave., Columbus, OH 43210, USA}
\affil[4]{Observatories of the Carnegie Institution of Washington, 813 Santa Barbara Street, Pasadena, CA 91101}

\def\corrauthor{
Andrew Gould
}

\def\runningauthor{
Gould \& Kollmeier

}

\def\runningtitle{
    Predicted Information Content of Gaia-Hipparcos
}

\def\keywords{
astrometry
}

\def\abstracttext{
The Gaia-Tycho release, scheduled for 14 September, is forecast to
yield parallax errors of $\sigma(\pi)\sim 300\,\mu$as for about
2 million Tycho stars.  We show analytically that the actual 
performance should be
$$
\sigma(\pi) = {\rm max}(\sigma_{1991}/96,20\,\mu{\rm as})
$$
where $\sigma_{1991}$ is the positional error from the Hipparcos
mission.  For typical Tycho stars, $\sigma_{1991}\sim 30\,$mas, so this
reproduces the usual claims.  However, for the 100,000 star Hipparcos subset
of this sample, $\sigma_{1991}$ is a factor 15 or more smaller.  These
much lower Hipparcos positional errors apply even to stars at
the Hipparcos-Tycho limit, $V\sim 12$.  This is especially important
for RR Lyrae stars, as well as other special classes, that were
systematically included in the Hipparcos catalog down to this limit
because of their exceptional scientific importance.  This predicted
performance will provide an early test of the Gaia algorithms.
}

\begin{document}
\jkashead 


\section{Introduction \label{sec:intro}}

On 14 September, Gaia will release parallax and proper motion measurements
for roughly 2 million stars in the Tycho catalog.  The release will
be based on about 9 months of Gaia data, which normally would not
be enough to disentangle the correlations between the three quantities,
position $P$, proper motion $\mu$, and parallax $\pi$.  However, for
Tycho stars, these degeneracies can be broken by the 1991 position
measurements of the Hipparcos satellite.  The ``anticipated'' parallax
precision (no doubt highly influenced by the fact that the actual
precisions are already known to the Gaia team) is $\sigma(\pi)\sim 300\,\muas$.
This release will enable a huge range of science, and also facilitate
community preparation for the subsequent full Gaia releases.

However, as we show here, the information content of these observations
is potentially much greater for the subset of these stars with Hipparcos
data.  These stars have much smaller positional errors than typical
Tycho stars, even when they are at similar $V$ magnitudes.  We ourselves
are motivated by the problem of calibrating RR Lyrae period-luminosity
relations, a project that will greatly benefit from the much higher
precision that we predict.  This calibration will also pave the way
to measuring the Gaia parallax zero point \citep{gk2016}.  However,
the same improved precision would greatly aid many other investigations
as well.

\section{{Analytic Gaia-Tycho Errors}
\label{sec:analytic}}

In order to elucidate the basic issues, we consider a highly
simplified case: three measurements of a star position, which lies in
the ecliptic, two of which are taken $2\delta t=6\,$months apart at
quadrature and the third is taken $\Delta t=24\,$yr previously.  Let
the errors for the two recent measurements both be $\sigma$, and let
the error in the early measurement be $\sigma_{1991}$.  For
simplicity, we assume that the mean epoch of the early measurement is
made at opposition.  Then there are three equations and three
unknowns:
\begin{equation}
P_{1991} = P - \mu\Delta t  \pm \sigma_{1991}
\label{eqn:1991}
\end{equation}
\begin{equation}
P_- = P -\mu\delta t -\pi \pm \sigma
\label{eqn:minus}
\end{equation}
\begin{equation}
P_+ = P +\mu\delta t +\pi \pm \sigma
\label{eqn:plus}
\end{equation}
where $P$ is the position at time $t=0$, i.e., halfway between the 
times of the two recent measurements.
Equations~(\ref{eqn:minus}) and (\ref{eqn:plus}) can be written more
compactly as $P_\pm = P \pm (\mu\delta t +\pi)$.
Using standard techniques (e.g. \citealt{gould03})
these equation yield the inverse covariance matrix $b_{ij}$
for the variables $(P,\mu,\pi)$ is
\begin{equation}
b_{ij}=\left(\matrix{w_{1991} + 2w & -w_{1991}\Delta t & 0\cr
-w_{1991}\Delta t & w_{1991}(\Delta t)^2 + 2w(\delta t)^2 & 2w\delta t\cr
0 & 2w\delta t & 2w}\right)
\label{eqn:bij}
\end{equation}
where $w\equiv \sigma^{-2}$ and $w_{1991}\equiv \sigma_{1991}^{-2}$ 
Inverting this matrix yields $c\equiv b^{-1}$, and so,
\begin{equation}
\sigma^2(\pi) = c_{33} = {\sigma^2\over 2}
\biggl(1 - {(w_{1991}+2w)(\delta t)^2\over 
2w(\delta t)^2 + w_{1991}((\Delta t)^2+(\delta t)^2)}\biggr)^{-1}.
\label{eqn:c33}
\end{equation}
That is, after some algebra,
\begin{equation}
\sigma^2(\pi) = \sigma_{1991}^2\biggl({\delta t\over\Delta t}\biggr)^2 +
{\sigma^2\over 2}\biggl[1+\biggl({\delta t\over\Delta t}\biggr)^2\biggr].
\label{eqn:sigpi}
\end{equation}
Since $(\delta t/\Delta t)^2\sim 10^{-4}$, Equation~(\ref{eqn:sigpi}) is
extremely well approximated by
\begin{equation}
\sigma(\pi) = \sqrt{\sigma_{1991}^2\biggl({\delta t\over\Delta t}\biggr)^2 +
{\sigma^2\over 2}}.
\label{eqn:sigpi2}
\end{equation}

The simplicity of Equation~(\ref{eqn:sigpi2}) relative to 
Equations~(\ref{eqn:bij}) and (\ref{eqn:c33}) implies that the
problem contains symmetries that are not captured by the above
brute-force analysis.  These can be understood as follows.  From the
two recent measurements, we have $P = (P_+ + P_-)/2\pm \sigma/\sqrt{2}$
and $(\mu\delta t+\pi) = (P_+ - P_-)/2\pm \sigma/\sqrt{2}$, with the
two quantities being completely uncorrelated.  This implies that
$\mu = (P-P_{1991})/\Delta t$ 
(with $[\sigma(\mu)\Delta t]^2=\sigma^2/2+\sigma_{1991}^2$) is also
uncorrelated with the $(\mu\delta t +\pi)$ measurement, just above.  Combining
these two therefore yields Equation~(\ref{eqn:sigpi}) (and so
Equation~(\ref{eqn:sigpi2})).

Now, if we assume that $\sigma_{1991}(\delta t/\Delta t)>\sigma$, then
Equation~(\ref{eqn:sigpi2}) simplifies to
\begin{equation}
\sigma(\pi) = \sigma_{1991}{\delta t\over \Delta t} = {\sigma_{1991}\over 96}
\qquad ({\rm large}\ \sigma_{1991}).
\label{eqn:sigpi3}
\end{equation}
Hence, for typical Tycho stars, with $\sigma_{1991}=30\,\mas$, the expected
parallax errors are $\sigma(\pi)\sim 300\,\muas$. i.e., exactly the
errors that are usually advertised for this catalog.

However, for stars in the Hipparcos catalog, $\sigma_{1991}$ is dramatically
smaller.  Even for $V\sim 12$ (the faintest stars that were routinely
targeted by Hipparcos, and also the effective limit of the Tycho catalog),
$\sigma_{1991}\sim 2\,\mas$, while for brighter stars it is typically
a factor few smaller.  Adopting this faint limit value, one finds
$(\delta t/\Delta t)\sigma_{1991}=20\,\muas$.  This is no longer large
compared to $\sigma$ and so we cannot simply apply Equation~(\ref{eqn:sigpi3}).

To make a rough estimate of what can be achieved for Hipparcos stars,
we must derive an effective value of $\sigma$.  First we note that
for the mission as a whole, the Gaia 
website\footnote{http://www.cosmos.esa.int/web/gaia/science-performance} gives 
\begin{equation}
\sigma_{\rm 5\,yr}(\pi) =(-1.631 + 680.766\, z + 32.732\, z^2)^{1/2}Q(V-I)
\label{eqn:gaiaperf}
\end{equation}
where $Q(V-I)\equiv [0.986 + 0.014(V-I)]$
and
\begin{equation}
z=\min(10^{0.4(G-15)},10^{-1.2}).
\label{eqn:gaiaperf2}
\end{equation}
Since essentially all Hipparcos stars have $G<12$, this equation simplifies
considerably to
\begin{equation}
\sigma_{\rm 5\,yr}(\pi) = 6.4\muas .
\label{eqn:gaiaperf3}
\end{equation}
Since the Gaia-Tycho release is based on just 9 months of data, we
adopt $\sigma/\sqrt{2}\rightarrow 6.4*\sqrt{60/9} = 16.5\,\muas$.
We discuss the role of this, and other simplifications we have made
in Section~\ref{sec:approx}, but the short answer is that none of these
affect our basic argument.

Therefore, for the general case
\begin{equation}
\sigma(\pi) = \sqrt{\biggl({\sigma_{1991}\over 96}\biggr)^2 + (16.5\,\muas)^2}.
\label{eqn:sigpi4}
\end{equation}
Hence, for the case of faint $(V\sim 12)$ Hipparcos stars with
$\sigma_{1991}=2\,\mas$, we have $\sigma(\pi) = 27\,\muas$.

\section{{Role of Approximations}
\label{sec:approx}}

In order to derive the results of Section~\ref{sec:analytic}, we
made a wide variety of approximations.  We now examine their potential
impact on these results.  

First, we approximated the Gaia measurements as consisting of two
observations at quadrature rather than a series of relatively random
observations constrained by visibility due to Sun-exclusion.  This
would seem to vastly oversimplify the correlations that the real-sequence
of measurements induces on the parallax and proper motion measurements.
However, the {\it sign} of this effect goes in the wrong direction
to give rise to concerns.  That is, a real series of measurements over
less than a year will induce some correlations -- possibly quite severe
-- between $\pi$ and $\mu$.  However, our simplification of two measurements
actually induced {\it perfect} correlation between these variables.  That is,
there is good information about $(\mu\delta t + \pi)$ but absolutely
zero information about $\mu$ and $\pi$ separately.  Thus, any real series
of measurements will actually contain {\it more} information to separate
$\pi$ and $\mu$ than we have assumed.

Second, we have assumed that the target lies on the ecliptic.  However,
this only causes us to underestimate the amount of information about
parallax.  On average the true errors for a given time sequence of
observations will be reduced by a factor $(1 + \sin^2\beta)^{-1/2}$ where
$\beta$ is the ecliptic latitude.  This correction is usually small,
but in any case only improves the parallax precision relative to our
estimates.

Third, we have estimated an effective $\sigma$ by simple root-N scaling
from the full mission, whereas the real leverage coming from 9 months
of data will vary significantly over the sky.  However, even if this
leads to factor two variation in this effective $\sigma$, it does not
alter the basic scaling of Equation~(\ref{eqn:sigpi3}) for the case
that $(\delta t/\Delta t)\sigma_{1991}$ is large compared to this value.
It only means that at some points on the sky, the limiting precision
will be 50\% lower than the $16.5\,\muas$ that we have estimated and
in other parts of the sky, it will be 50\% higher.  The actual values
of these errors for particular stars will be important for any particular
application of the Gaia-Tycho catalog, but these variations do not
impact the basic argument given here.

Fourth, we have assumed that the 2015 Gaia and the 1991 Hipparcos (or Tycho)
positions are in the same global reference frame, whereas in fact
these frames are established independently.  Moreover, whereas
Hipparcos (like Gaia) parallaxes are absolute, its positions and 
proper motions had to be aligned to a radio-quasar based reference
frame via a handful of radio stars.  While the zero-point position
error plays no role in the current study (because it induces errors
that are an order of magnitude smaller than other inputs), the
proper motion reference frame error can be significant.

Recall from the ``heuristic'' derivation of Equation~(\ref{eqn:sigpi2})
(which is mathematically identical to the rigorous derivation),
that the proper motion is determined with precision 
$\sigma(\mu)=\sigma_{1991}/\Delta t$ (for the case of large $\sigma_{1991}$).
The error in $\pi$ is then $\sigma(\pi) = \sigma(\mu)\delta t$.
However, if the reference frame itself has a systematic error $\delta\mu$,
then this sets a floor $\sigma(\mu)\geq \delta\mu$, and hence a floor
$\sigma\pi\geq\delta\mu\delta t\rightarrow 60\,\muas$, where we have
adopted $\delta\mu=0.25\,\masyr$.  This is still quite good, and in
particular much better than the $300\,\muas$ errors for Tycho stars.

In fact, however, it should be possible to recover the full power of
Equation~(\ref{eqn:sigpi2}).  The proper-motion reference frame consists
of three numbers, i.e., $\bOmega$, which give the rate of rotation
of the reference frame relative to the ``true'' quasar frame.
Consider first Hipparcos stars near the ecliptic.
Each, of course, has two proper motion components, one of which
can be taken to perpendicular to the ecliptic (and the
other parallel).  The first, by construction, is completely
uncorrelated with the parallax, and so can be determined with
high precision $(\sim 66\,\muas\,\yr^{-1})$ 
from Gaia data alone.  This can then be directly compared
to the Gaia-Hipparcos proper motion for several $10^4$ stars, which
will yield the zero point of the Hipparcos proper motion frame for two
of the three components to $(<1\,\muas\,\yr^{-1})$.  
Next, we consider stars at moderate
ecliptic latitude, for which the parallax and proper motions are
coupled in both directions.  For an individual star, one can then
determine the parallax from the parallactic motion in the direction
with well-determined proper-motion zero point,  Then, from this
known parallax, one can determine the proper motion from Gaia-only
data in the other direction (with somewhat larger parallactic motion),
and so determine the proper motion in that direction, thereby
calibrating the Hipparcos zero point in that direction as well  Actually,
this argument is made only to illustrate the information content
of the data.  In practice, one would simultaneously fit for the
3 Hipparcos proper-motion zero points $\bOmega$ together with all other 
parameters.

Finally, we have assumed that what Gaia measures are positions on the
sky, whereas in fact the actual measurements are of relative separations
of stars near the ``basic angle'' $\sim 106^\circ$.  All astrometric
parameters are then derived by simultaneously solving a very large number
of equations derived from these offset measurements.  Within this same
process, one must also solve for a large number of parameters that describe
the spacecraft, including its orientation, small changes in the basic angle,
and many others. Nothing of this process is in any way captured by our
simple ``three measurement'' idealization.   Moreover, the Gaia software
for extracting these parameters (both science and engineering), was designed
to operate on about $10^9$ stars, whereas the Gaia-Tycho solutions are
based on only $10^6$ stars, of which only $10^5$ have high-precision,
Hipparcos, 1991-position measurements.

However, while we do not claim to be experts in these data reductions, the basic
answer to these concerns is that the information flow for Gaia-Tycho
(or, really, Gaia-Hipparcos) is nearly the same as for Hipparcos:
same number of stars, same astrometric parameters, and basically
same spacecraft parameters.  The errors are smaller, but the mathematical
treatment simply scales with these.  The data stream is shorter,
so there are
different types of -- and stronger -- correlations between parameters
from Gaia-only (compared to Hipparcos-only) measurements.  However,
because ``Gaia-only'' measurements are supplemented by first-epoch
measurements from Hipparcos, the actual correlations between parameters
are similarly small for Gaia+Hipparcos as for Hipparcos.

\section{{Discussion}
\label{sec:discuss}}

It may well be that when the Gaia-Tycho catalog is released, the
reported errors will be roughly as predicted here.  In this case,
the main value of the present work will be to alert the community
to the fact that much higher-precision science is possible than
was thought based on pre-release advertisements.  For example,
typical Hipparcos RR Lyrae stars have parallaxes $\pi\sim 800\,\muas$
(e.g., \citealt{popow98b}).  If the parallax measurements have
errors of $\sigma(\pi)\sim 300\,\muas$, then it will be possible
to determine the zero-point of the period-luminosity (PL) relation to
a precision of $(5/\ln 10)(300/800)/\sqrt{100}= 0.08\,$mag from a sample
of 100 stars.  This is really not qualitatively better than a number of previous
determinations dating back more than 20 years 
\citep{longmore90,layden96,popow98a,gould98,benedict11,kollmeier13,madore13,dambis14}.  On the other hand, if the precisions can be improved a factor 10, then
the PL zero points will experience similar improvements.  Moreover,
such improvements would allow qualitatively different questions can be 
addressed.  For example, the scatter about the PL relation could be 
investigated in field RR Lyrae stars, whereas currently this is only
possible in (presumably more homogeneous) clusters.

On the other hand, if the errors reported in the Gaia-Tycho catalog
are at the $300\,\muas$ level even for Hipparcos stars, then the
analytic arguments presented here can be used to track down the
discrepancy between these reports and our predictions.  This may lead
either to an improved catalog or to a deeper understanding of issues
that will impact future Gaia releases.


\acknowledgments
We thank Amelia Stutz for valuable comments.
This work was supported by NSF grant AST-1516842.


\end{document}